\documentclass{article}
\usepackage{graphicx} 
\usepackage{amsmath}  
\usepackage{amssymb}  
\usepackage{geometry} 
\usepackage{authblk}  
\usepackage{setspace} 
\usepackage{booktabs}
\usepackage{multirow}
\usepackage{threeparttable}
\usepackage{caption}
\usepackage{geometry}
\usepackage{array}
\usepackage{longtable}
\usepackage{natbib} 
\title{Genetic association testing with multivariate survival phenotypes under interval censoring}
\author[1]{Juhee Lee}
\author[2]{Kun Xia}
\author[2]{Jianrui Zhang}
\author[3]{Gongjun Xu}
\author[4]{Qing Lu}
\author[2]{Chenxi Li}
\affil[1]{Department of Statistics and Probability, Michigan State University, East Lansing, MI, USA}
\affil[2]{Department of Epidemiology and Biostatistics, Michigan State University, East Lansing, MI, USA}
\affil[3]{Department of Statistics, University of Michigan, Ann Arbor, MI, USA}
\affil[4]{Department of Biostatistics, University of Florida, Gainesville, FL, USA}
\date{}

\begin{document}

\maketitle

\begin{abstract}
Set-based genetic association tests provide a powerful framework for detecting genetic effects on complex traits by jointly analyzing multiple genetic variants. Although set-based methods have been developed for interval-censored survival outcomes, existing approaches primarily focus on a single survival phenotype and therefore do not fully use information from multiple correlated outcomes. In this paper, we develop two Weighted V Tests for Multivariate Interval-Censored Data (WV-M-IC), extending the weighted V-statistic framework (Wu et al., 2021) to the joint analysis of multiple correlated interval-censored survival outcomes. The performance of these methods is evaluated through simulation studies, showing that the proposed approaches can provide power gains compared with single-outcome analyses. We apply the proposed methods to the ZOE 2.0 study to investigate dental caries progression in children.
\end{abstract}

\section{Introduction}

Genetic association studies are widely used to investigate the genetic basis of complex traits and diseases. Advances in genotyping and sequencing technologies have made it possible to study genetic variation across the genome at a large scale. However, individual genetic variants often have small effects, making it difficult to detect genetic associations through single-marker analyses. In addition, testing a large number of variants individually introduces a substantial multiple-testing burden. These limitations have motivated the development of set-based association tests, which jointly evaluate multiple genetic variants within a biologically meaningful unit, such as a gene or genomic region. By aggregating genetic signals across variants, set-based approaches can improve power to detect genetic associations that may be difficult to identify through individual-marker analyses. Among these methods, kernel-based association tests jointly evaluate multiple genetic variants and can accommodate complex genetic effects through kernel functions \citep{liu2008estimation,liu2007semiparametric,wu2011rare}.

Set-based association methods have also been developed for survival outcomes, which provide information on the timing of disease onset or progression and are therefore useful for studying genetic effects on disease development. Several kernel-based and variance-component approaches have been proposed for genetic association analysis of censored survival outcomes. Among these methods, the weighted V test (WV) proposed by \cite{li2021set} provides a flexible framework for detecting complex genetic effects, such as nonlinear or interactive effects, of a SNP set or gene set on survival outcomes. \citet{li2021set} demonstrated empirically that WV maintains the nominal type I error rate in moderate-sized samples and showed both theoretically and empirically that WV with a linear kernel maintains the nominal type I error rate when adjusting for covariates that are linearly correlated with the genetic markers of interest. Furthermore, WV can accommodate left truncation of survival traits and account for genetic effect heterogeneity, which may increase power in the presence of heterogeneous genetic effects. These properties make the weighted V framework a useful basis for extending set-based association methods to more complex survival data structures.

An additional challenge arises when survival outcomes are interval-censored. In many biomedical and epidemiological studies, particularly those involving chronic diseases, disease status is assessed intermittently rather than continuously. As a result, the exact time of disease onset is unknown, and the event time is only known to fall within an interval defined by two observation times. For example, the onset of early childhood caries (ECC) may be interval-censored when disease status is evaluated only at periodic visits. Methods developed for right-censored survival outcomes do not directly account for the information contained in such interval-censored observations. To address this issue, \cite{wu2021multi} developed the weighted V tests for univariate interval-censored survival outcomes (WV-IC), extending the weighted V framework to interval-censored data. The WV-IC framework provides set-based genetic association and interaction tests for interval-censored survival outcomes while retaining the weighted V framework. However, these methods are formulated for a single survival phenotype and therefore do not directly use the information shared across multiple correlated interval-censored outcomes.

Certain genetic variants may affect multiple related diseases or phenotypes through shared biological mechanisms \citep{li2022multi}. Therefore, jointly analyzing multiple related outcomes can provide additional information and potentially improve the detection of genetic associations. \cite{choi2024variance} proposed a set-based variance-component test for genetic association with multiple interval-censored outcomes and demonstrated that jointly analyzing multiple outcomes can improve power for detecting rare and weak genetic effects. Their approach provides a useful framework for jointly analyzing multiple interval-censored outcomes. However, it is developed within a variance-component testing framework, while the weighted V framework provides a flexible approach for detecting complex genetic effects, including nonlinear and interactive effects. Motivated by the weighted V framework, we extend it to the multivariate interval-censored setting to jointly analyze correlated survival phenotypes.

We develop two Weighted V Tests for Multivariate Interval-Censored Data (WV-M-IC), a set-based framework for genetic association analysis of multiple correlated interval-censored survival phenotypes. The proposed methods extend the weighted V framework to jointly incorporate information from multiple outcomes while accounting for their correlation structure. By using information across outcomes, the proposed tests can improve power relative to analyses based on individual phenotypes, particularly when genetic effects are shared across correlated outcomes, while maintaining appropriate type I error control. We evaluate the finite-sample performance of the proposed methods through extensive simulation studies under varying outcome correlation structures and genetic effect settings. We further demonstrate the practical utility of the proposed methods through a gene-based analysis of early childhood caries using data from the ZOE 2.0 study \citep{ginnis2019measurement,divaris2020cohort}.

\section{Methods}

\subsection{Set-up}
We aim to test the effect of a marker set $\mathbf{G}=(G_1,\ldots,G_p)^T$ on multiple failure times $T_1, \ldots, T_m$ (e.g., ages to caries on multiple primary teeth) simultaneously, adjusting for a set of covariates $\mathbf{Z}=(Z_1,\ldots,Z_q)^T$. Integrating more information from a multivariate survival outcome as opposed to a univariate outcome is expected to lead to a higher power of the association test. $\mathbf{G}$ can be a set of SNPs, gene expressions, epigenetic markers, etc. The marker sets can be formed based on genes, pathways, LD blocks, recombination hot-spots, or using a sliding window approach. $\mathbf{Z}$ are confounders and/or predictors independent of $\mathbf{G}$. Under interval censoring, the observed survival data are $\{L_{ik},R_{ik}\}^{n,m}_{i=1,k=1}$, where $(L_{ik},R_{ik})$ is the bracketing interval of subject $i$'s $T_k$.

\subsection{WV-M-IC Test Statistic 1}
We assume that $T_k$ ($k=1,\dots,m$) marginally follows a semiparametric transformation model \citep{zeng2016maximum} under the null hypothesis that $\mathbf{G}$ is independent of $(T_1,\dots,T_m)$ given $\mathbf{Z}$. Define
\[
S_{\mathbf{Z},ik} = \left. \frac{\partial \log L_{ik}(\Lambda_{0k}, \boldsymbol{\gamma}_k, h_{ik})}{\partial h_{ik}} \right|_{h_{ik}=0},
\]
where $L_{ik}(\Lambda_{0k}, \boldsymbol{\gamma}_k, h_{ik}) = \exp\left(-G\left[\Lambda_{0k}(L_{ik}) e^{\boldsymbol{\gamma}_k' \mathbf{Z}_i + h_{ik}}\right]\right) - \exp\left(-G\left[\Lambda_{0k}(R_{ik}) e^{\boldsymbol{\gamma}_k' \mathbf{Z}_i + h_{ik}}\right]\right)$
is the conditional likelihood (given frailty $e^{h_{ik}}$) of subject $i$'s ($i=1,\dots,n$) data based on a semiparametric transformation frailty model $\Lambda_k(t|\mathbf{Z}, h_k) = G\left[\Lambda_{0k}(t) e^{\boldsymbol{\gamma}_k' \mathbf{Z} + h_k}\right],$
$\Lambda_k(t|\mathbf{Z}, h_k)$ is the cumulative hazard function of $T_k$ given $\mathbf{Z}$ and $h_k$, $G(x) = \frac{\log(1 + rx)}{r}$ with a specified $r$ ($r \geq 0$), and $\Lambda_{0k}(t)$ is an unknown nondecreasing function with $\Lambda_{0k}(0) = 0$. Let $\vec{S}_{\mathbf{Z},i}^T = (S_{\mathbf{Z},i1}, \dots, S_{\mathbf{Z},im})$. The proposed test is based on a weighted $V$ statistic,
\[
V_{\mathbf{Z},M-IC}^{*(1)} = n^{-2} \sum_{i=1}^n \sum_{j=1}^n \tilde{f}_\mathbf{Z}(\mathbf{G}_i, \mathbf{G}_j) \vec{S}_{\mathbf{Z},i}^T \Sigma_{\vec{S}_\mathbf{Z}}^{-1} \vec{S}_{\mathbf{Z},j}, \tag{1}
\]
where $\Sigma_{\vec{S}_\mathbf{Z}} = \text{Cov}(\vec{S}_{\mathbf{Z},1})$. $\vec{S}_{\mathbf{Z},i}^T \Sigma_{\vec{S}_\mathbf{Z}}^{-1} \vec{S}_{\mathbf{Z},j}$ is considered as the $V$-statistic kernel and is a covariate-adjusted phenotype similarity measure. $\tilde{f}_\mathbf{Z}(\mathbf{G}_i, \mathbf{G}_j)$ is considered as the weight function and is a covariate-centered genetic similarity measure defined by
\begin{align*}
\tilde{f}_\mathbf{Z}(\mathbf{G}_i, \mathbf{G}_j) &= f(\mathbf{G}_i, \mathbf{G}_j)
+ \mathbb{E}\left[u(\mathbf{Z}_i, \mathbf{Z}_m)f(\mathbf{G}_m, \mathbf{G}_k)u(\mathbf{Z}_k, \mathbf{Z}_j) \mid \mathbf{Z}_i, \mathbf{Z}_j\right]\\
&- \mathbb{E}\left[f(\mathbf{G}_i, \mathbf{G}_k)u(\mathbf{Z}_k, \mathbf{Z}_j) \mid \mathbf{G}_i, \mathbf{Z}_j\right]
- \mathbb{E}\left[u(\mathbf{Z}_i, \mathbf{Z}_k)f(\mathbf{G}_k, \mathbf{G}_j) \mid \mathbf{Z}_i, \mathbf{G}_j\right] \\
\end{align*}
where $f(\mathbf{G}_i, \mathbf{G}_j)$ is a genetic similarity function and $u(\mathbf{Z}_i, \mathbf{Z}_j) = (1, \mathbf{Z}_i^T) \left[\mathbb{E}\{(1, \mathbf{Z}^T)^T (1, \mathbf{Z}^T)\}\right]^{-1} (1, \mathbf{Z}_j^T)^T.$
It can be shown that $\mathbb{E}\left[\tilde{f}_\mathbf{Z}(\mathbf{G}_i, \mathbf{G}_j)(1, \mathbf{Z}_j^T) \mid \mathbf{G}_i, \mathbf{Z}_i\right] = 0$. The choice of $f(\mathbf{G}_i, \mathbf{G}_j)$ depends on the type of $\mathbf{G}$ and the expected form of $\mathbf{G}$'s effect. If the effect of $\mathbf{G}$ is expected to be linear, we use the linear kernel, $f(\mathbf{G}_i, \mathbf{G}_j) = \mathbf{G}_i^T \mathbf{G}_j.$
For SNPs, we use the IBS kernel, $f(\mathbf{G}_i, \mathbf{G}_j) = \sum_{k=1}^p \frac{2 - |\mathbf{G}_{i,k} - \mathbf{G}_{j,k}|}{2p},$
if the effect of $G$ is not expected to be linear. For gene expression levels, we use the polynomial kernel $f(\mathbf{G}_i, \mathbf{G}_j) = \left(\rho + \mathbf{G}_i^T \mathbf{G}_j\right)^d$
or the Gaussian kernel $f(\mathbf{G}_i, \mathbf{G}_j) = \exp\left(-\rho \|\mathbf{G}_i - \mathbf{G}_j\|^2\right)$
if nonlinear or interactive effects of $\mathbf{G}$ are expected. As suggested by \cite{wei2017generalized}, we can also choose a unified Laplacian kernel based generic similarity, $f(\mathbf{G}_i, \mathbf{G}_j) = \exp\left(-\sum_{k=1}^p \frac{w_k |\mathbf{G}_{i,k} - \mathbf{G}_{j,k}|}{\Upsilon}\right)$, 
where $\mathbf{G}_{i,k}$ can be discrete or continuous variables, $\Upsilon = \sum_{k=1}^p w_k$, and $w_k$ is a function of variance $\sigma_k^2$ of $\mathbf{G}_k$ defined by $w_k = \frac{1}{\sigma_k}$. This kernel is particularly useful for association analyses with sequencing data, where a large portion of genetic variants are rare.

Under the null hypothesis the $\mathbf{G}$ has no effect on any $T_k$ $(k=1,\ldots, m)$ adjusting for $\mathbf{Z}$, $E(S_{\mathbf{Z},ik}|\mathbf{Z}_i)=0$. When $\mathbf{Z}$ is independent of $\mathbf{G}$, following Theorem 1 in \cite{li2021set}, the null limiting distribution of $nV_{\mathbf{Z},M-IC}^{*(1)}$ can be shown to be \(\sum_{t=1}^\infty \nu_t \chi_{mt}^2,\) where $\chi^2_{mt}$ are independent chi-square variates with degree $m$ and $v_t$'s are the eigenvalues of $\tilde{f}(\mathbf{g}_i,\mathbf{g}_j) = \sum^\infty_{t=1} v_t\phi_t(\boldsymbol{g}_i,\boldsymbol{z}_i)\phi_t(\boldsymbol{g}_j,\boldsymbol{z}_j)$ with \( E(\phi_s(\mathbf{G}, \mathbf{Z}) \phi_t(\mathbf{G}, \mathbf{Z})) = I(s = t) \). When \( \mathbf{Z} \) affects both \( \mathbf{G} \) and the survival outcome, the test may be slightly conservative based on the discussion of \cite{li2021set}. However, when the linear kernel is used for \( f(\mathbf{G}_i, \mathbf{G}_j) \), the test is still asymptotically correct under linear confounding, i.e., \( \mathbf{G} = \mathbf{a} + \mathbf{B}^T \mathbf{Z} + \mathbf{e} \), where \( \mathbf{a} \) and \( \mathbf{B} \) are respectively a constant vector and a constant matrix, and \( \mathbf{e} \) is a zero-mean random error that is independent of $\mathbf{Z}$. It can be shown, following the proof of Theorem 2 in \cite{li2021set}, that the null limiting distribution of \( n V_{\mathbf{Z}, M-IC}^{*(1)} \) with the linear kernel is (1) under linear confounding. Under the alternative hypothesis that \( \mathbf{G} \) has effects on the survival outcome adjusting for \( \mathbf{Z} \), the covariate-adjusted phenotype similarity \( \vec{S}_{\mathbf{Z},i}^T \Sigma_{\vec{S}_\mathbf{Z}}^{-1} \vec{S}_{\mathbf{Z},j} \) is concordant with the covariate-centered genetic similarity \( \tilde{f}_\mathbf{Z}(\mathbf{G}_i, \mathbf{G}_j) \). In other words, the larger phenotype similarity is weighted heavier and the smaller phenotype similarity is weighted lighter, leading to a large value of \( V_{\mathbf{Z}, M-IC}^{*(1)} \).

The vector \( \vec{S}_{\mathbf{Z},i}^T \) involves unknown parameters \( (\Lambda_{0k}, \gamma_k) \) (for \( k = 1, \ldots, m \)), the covariance matrix \( \Sigma_{\vec{S}_\mathbf{Z}} \) is unknown, and \( \tilde{f}_\mathbf{Z}(\mathbf{G}_i, \mathbf{G}_j) \) involves unknown conditional expectations. In real applications, we replace $(\Lambda_{0k},\gamma_k)$ by the estimates $(\hat{\Lambda}_{0k},\hat{\gamma}_k)$ obtained from a nonparametric maximum likelihood estimation for the model $\Lambda_k(t|\mathbf{Z},h=0)=G[\Lambda_{0k}(t)e^{\gamma'_k\mathbf{Z}}]$. The resulting $\vec{S}_{\mathbf{Z},i}^T$ is denoted by $\hat{\vec{S}}_{\mathbf{Z},i}^T$. $\Sigma_{\vec{S}_{\mathbf{Z}}}$ will be replaced by an empirical estimate $\hat{\Sigma}_{\vec{S}_{\mathbf{Z}}}=n^{-1}\sum^n_{i=1}\hat{\vec{S}}_{\mathbf{Z},i}\hat{\vec{S}}_{\mathbf{Z},i}^T$. We replace the expectations in $\tilde{f}_\mathbf{Z}(\mathbf{G}_i,\mathbf{G}_j)$ by the corresponding sample averages. The resulting $\tilde{f}_\mathbf{Z}(\mathbf{G}_i,\mathbf{G}_j)$ equals the $ij$-th element of the matrix $(I-H)^TF(I-H)$, where $F=\{\tilde{f}(\mathbf{G}_i,\mathbf{G}_j)\}_{n \times n}$, $I$ is an $n \times n$ identity matrix, and $H=\tilde{Z}(\tilde{Z}^T\tilde{Z})^{-1}\tilde{Z}^T$ with $\tilde{Z}$ being a $n \times (q+1)$ matrix whose $i$-th row is $(1,\mathbf{Z}_i^T)$ $(i=1,\ldots,n)$. The corresponding weighted V statistics is denoted by $V_{\mathbf{Z},M-IC}^{(1)}$. We use a matrix eigen-decomposition of $(I-H)^TF(I-H)$ to obtain a finite-dimensional approximation for $v_t$'s. Denote the eigen-values and eigen-vectors of $(I-H)^TF(I-H)$ by $\tilde{v}_t$ $(t=1,\ldots,n)$ respectively. Because $\tilde{\phi}_t$ satisfies $\sum^n_{i=1}\tilde{\phi}^2_{t,i}=1$, instead of $n^{-1}\sum^n_{i=1}\tilde{\phi}^2_{t,i}=1$, a finite-dimensional approximation for $v_t$ is $\hat{v}_t=\tilde{v}_t/n$. Then the large sample null distribution of $nV_{\mathbf{Z},M-IC}^{(1)}$ is approximately
\[
nV_{\mathbf{Z},M-IC}^{(1)}={1 \over n} \sum^n_{i=1}\sum^n_{j=1} \tilde{f}(\mathbf{G}_i,\mathbf{G}_j)\vec{S}_{\mathbf{Z},i}^T\Sigma_{\vec{S}_\mathbf{Z}}^{-1}\vec{S}_{\mathbf{Z},j} \sim \sum^n_{t=1} \hat{v}_t\chi^2_{mt}. \tag{2}
\]
Based on this distribution, we use Davies' method \citep{davies1980algorithm} to compute the p-value $P(nV_{\mathbf{Z},M-IC}^{(1)} \ge nV_{\mathbf{Z},M-IC,obs}^{(1)})$.

\subsection{WV-M-IC Test Statistic 2}
Similarly, the second multivariate WV test is based on a weighted $V$ statistic,
\[
V_{\mathbf{Z},M-IC}^{*(2)} = n^{-2} \sum_{i=1}^n \sum_{j=1}^n \tilde{f}_\mathbf{Z}(\mathbf{G}_i, \mathbf{G}_j) \vec{S}_{\mathbf{Z},i}^T \vec{S}_{\mathbf{Z},j}. \tag{3}
\]
Compared with the first test, the second test does not use \(\Sigma_{\vec{S}_\mathbf{Z}}^{-1}\) in the phenotype similarity measure. The large sample null distribution of \(nV_{\mathbf{Z},M-IC}^{(2)}\) is approximately
\[
nV_{\mathbf{Z},M-IC}^{(2)}={1 \over n} \sum^n_{i=1}\sum^n_{j=1} \tilde{f}(\mathbf{G}_i,\mathbf{G}_j)\vec{S}_{\mathbf{Z},i}^T\vec{S}_{\mathbf{Z},j} \sim \sum^n_{t=1} \hat{v}_t\left(\sum^m_{k=1}\lambda_k \chi^2_{kt}\right), \tag{4}
\]
where $\lambda_1,\ldots,\lambda_m$ are the eigenvalues of $\Sigma_{\vec{S}_\mathbf{Z}}$. Based on this distribution, we use Davies' method \citep{davies1980algorithm} to compute the p-value $P(nV_{\mathbf{Z},M-IC}^{(2)} \geq nV_{\mathbf{Z},M-IC,obs}^{(2)})$.

\section{Simulations}
We conducted Monte Carlo simulations to evaluate the finite-sample performance of the proposed weighted V tests for multivariate interval-censored outcomes. In all simulation scenarios, subjects' baseline covariates were adjusted, and survival times were generated from proportional hazards ($r=0$) and proportional odds ($r=1$) models, representing two special cases of semiparametric transformation models. For each subject, three follow-up times, $V_1, V_2, V_3$, were simulated to control censoring rates, with $V_1 \sim \text{Unif}(0,1)$, $V_2 \sim V_1 + \text{Unif}(0,2)$, and $V_3 \sim V_2 + \text{Unif}(0,2)$. Left and right censoring rates were controlled within 15\%-25\% and 35\%-45\%, respectively. We examined two sample sizes, $n=400$ and $n=800$, and two marker-set sizes, $p=15$ and $p=25$, to assess performance across different data scales. Adjustment covariates included a binary covariate $Z_1$ and a continuous covariate $Z_2 \sim \text{Unif}(0,2)$. Each scenario comprised 1000 Monte Carlo replicates, with a nominal significance level of 0.05.

We also considered combining the results from single-outcome tests as an alternative approach for analyzing multiple interval-censored outcomes. Specifically, we applied the univariate weighted V test (WV-IC) \citep{wu2021multi} separately to each outcome and combined the resulting $p$-values using the Bonferroni correction \citep{bland1995multiple}, Simes test \citep{simes1986improved}, or Aggregated Cauchy Association Test (ACAT) \citep{liu2019acat}.

\subsection{Empirical size and power under various n's and p's}
To evaluate the empirical size and power of WV-M-IC, we considered eight failure times corresponding to the eight primary molar teeth (T1-T8). We generated eight survival times for each subject $i$ ($i=1,\ldots,n$) using the inverse CDF method under a proportional hazards model with the following hazard function,
\[
\lambda(t|\boldsymbol{G},\boldsymbol{Z}) = \exp\{\sum_{j=1}^{p}\beta_{1j}G_{ij} + \sum_{k=1}^{2}\beta_{2k}Z_{ik} \} 
\]
where $t$ is the survival time, $G_{ij}$ are genetic markers, and $Z_{ik}$ represents covariates. The correlation among 8 failure times was induced by a Gaussian copula with a correlation matrix having off-diagonal elements $\rho$ ($\rho=0.1, 0.25, 0.5, 0.75$). The coefficients $\beta_{1j}$ and $\beta_{2k}$ were specified according to the simulation scenario. For all scenarios, we fixed $\beta_{2k}=0.05$ for $k=1,2$. For the size assessment, we set $\beta_{1j}= 0$ to simulate the null hypothesis of no genetic effect. For power assessment, we introduced non-zero genetic effects to mimic realistic scenarios. In the first scenario, we modeled the first four survival outcomes corresponding to the molars with the largest genetic effects, with coefficients set as $\beta_{1j}=(0.02,0.02,0.015,0.015)$, assuming that symmetrically positioned molars (left and right) share the same genetic effect. In the second scenario, we included all eight survival outcomes with coefficients defined as $\beta_{1j}=(0.02,0.02,0.015,0.015,0.01,0.01,0.005,0.005)$, reflecting paired effects across the four molar positions. The IBS kernel was applied to measure genetic similarity between subjects in WV-M-IC. Tables 1-8 show that the empirical sizes of WV-M-IC were close to 0.05 across all simulations. The results indicate that no single test consistently outperformed the others across all scenarios. Specifically, WV-M-IC1 exhibited higher power when genetic effects were present across multiple outcomes and the correlation between outcomes was high (e.g., $\rho \geq 0.75$). In contrast, WV-M-IC2 showed relatively stable power across different correlation structures and genetic-effect patterns, including scenarios in which genetic effects were present only for a subset of outcomes.

\subsection{Empirical Sizes of WV-M-IC Tests Under Stringent $p$-Value Thresholds}
Because genome-wide association studies require stringent significance thresholds to control false-positive findings across a large number of genetic tests, we further evaluated the empirical sizes of the WV-M-IC tests under small significance levels. We used the same simulation setting as in Scenario 2 of the previous size assessment, which included eight correlated interval-censored survival outcomes. Genetic effects were set to zero ($\beta_{1j}=0$) to represent the null hypothesis of no genetic association. We generated 150,000 Monte Carlo replicates for each combination of $n=400,800$ and $p=15,25$, allowing more precise evaluation of type I error rates under stringent significance thresholds.

Tables 9 and 10 show that the empirical sizes of the WV-M-IC tests remained close to the nominal significance levels across the considered thresholds. These results indicate that the proposed tests maintain appropriate type I error control even under stringent $p$-value thresholds relevant to large-scale genetic association studies.

\section{Application to a dental caries dataset}

Dental caries is the most common chronic childhood disease worldwide. Early childhood caries (ECC), which affects the primary dentition, is a particularly important oral health problem because caries experience in early childhood can have long-term consequences for oral health. In this study, we performed a genome-wide association analysis of ECC using the proposed multi-marker association tests. The dataset was derived from the ZOE 2.0 study, a genetic epidemiology study of early childhood caries \citep{ginnis2019measurement,divaris2020cohort}. The analysis included participants who were younger than 6 years at the time of their dental examination. ECC was assessed for the eight primary molars (A, B, I, J, K, L, S, and T). We used Trait6 as the primary phenotype definition, which includes established/moderate carious lesions beginning at ICDAS $\geq 3$ and excludes early-stage lesions (ICDAS 1 and 2). For each molar, the phenotype was defined as the age at ECC onset, which was subject to interval censoring because the exact time of caries onset was not directly observed.

SNP-level and subject-level quality control of the array data was conducted using PLINK 2.0. Specifically, a SNP was removed if any of the following criteria was met: (1) MAF $<0.01$, (2) HWE test $p$-value $<10^{-6}$, or (3) missing rate $>5\%$. A subject was removed if his or her genotype missing rate was greater than 5\%. After quality control, the directly genotyped array dataset contained 5,805 subjects and 888,057 SNPs. 
After matching the merged genotype data with the phenotype and covariate data, 5,587 subjects remained in the final analysis cohort.

For the gene-based analyses, the genotype data were represented by 23,210 genomic regions containing 7,001,566 SNP entries. The genomic regions were defined based on the hg38 reference genome. The 23,210 regions were then used as the units of gene-based association testing. For each region, the association with age at ECC onset across the eight primary molars was evaluated using the proposed weighted V tests for multivariate interval-censored outcomes (WV-M-IC1 and WV-M-IC2). In both methods, the eight molar-specific interval-censored outcomes were analyzed jointly, allowing the proposed tests to capture multivariate genetic associations across the primary molars. Genetic similarity among subjects was quantified using the identity-by-state (IBS) kernel. The first eight principal components (PCs) of the genotype data were included as ancestry covariates.

For comparison, we also evaluated the weighted V test for interval-censored outcomes (WV-IC). Unlike WV-M-IC1 and WV-M-IC2, which directly provide a single gene-level $p$-value by jointly analyzing the eight molar-specific outcomes, WV-IC was applied separately to each molar and therefore produced eight marginal $p$-values for each genomic region. To obtain a single region-level $p$-value from these marginal tests, we considered three commonly used $p$-value combination approaches: Bonferroni, Simes, and the aggregated Cauchy association test (ACAT). Thus, the comparison included five approaches: WV-M-IC1, WV-M-IC2, Bonferroni-combined WV-IC, Simes-combined WV-IC, and ACAT-combined WV-IC. All association analyses were adjusted for sex, race/ethnicity, and the top eight principal components of the genotype data to account for potential confounding due to population structure. Race/ethnicity was categorized into six groups: African American, American Indian/Alaskan Native, Asian, Native Hawaiian or Other Pacific Islander, White, and more than one race.

In all analyses, we considered two semiparametric transformation models: the proportional odds model ($r=1$) and the proportional hazards model ($r=0$). For the multivariate analyses, WV-M-IC1 and WV-M-IC2 were evaluated under each transformation model. The corresponding WV-IC analyses were likewise conducted under both transformation models before applying the Bonferroni, Simes, or ACAT $p$-value combination procedures. The false discovery rate (FDR) was controlled at 10\% using the Benjamini-Hochberg procedure \citep{benjamini1995controlling} without the assumption that tests were independent of each other. The corresponding $p$-value threshold is $l_i = (i\alpha) \left\{ m \sum_{k=1}^m (1/k) \right\}^{-1},$ for $i=1,\ldots,m$, where $m = 23,210$ and $\alpha$, the target FDR level, is 10\%.

The real-data analysis was performed in R 4.3.2 on the High Performance Computing Center (HPCC) cluster. To facilitate large-scale computation, the 23,210 genomic regions were divided into 465 parallel tasks, with each task testing up to 50 regions and the final task testing 10 regions. The 465 tasks were executed in parallel using the multi-node computing resources of the HPCC. The average computation time was approximately 3 days per task.

The association analysis results are summarized in Tables 11-14. No genomic region reached the multiple-testing-adjusted significance threshold under either the WV-M-IC1 or WV-M-IC2 analyses. Similarly, no region reached the significance threshold after combining the molar-specific WV-IC $p$-values using Bonferroni, Simes, or ACAT. Nevertheless, comparison of the top-ranked regions provides insight into the differences between the proposed multivariate tests and conventional $p$-value combination approaches. The Bonferroni, Simes, and ACAT approaches combine marginal association evidence from the eight molars, whereas WV-M-IC1 and WV-M-IC2 jointly incorporate information from the eight molar-specific outcomes within a single association test. The resulting rankings provide a comparison of how joint multivariate evidence and marginal $p$-value combination approaches prioritize candidate regions. The rankings were also compared across the proportional hazards ($r=0$) and proportional odds ($r=1$) models to assess the robustness of the results to the choice of transformation model.

The association analysis results are summarized in Tables 11--14. No genomic region reached the multiple-testing-adjusted significance threshold under either WV-M-IC1 or WV-M-IC2. Similarly, no region reached the significance threshold after combining the molar-specific WV-IC $p$-values using Bonferroni, Simes, or ACAT. Nevertheless, we examined the top-ranked regions to compare the signals identified by the proposed multivariate tests with those identified by the conventional $p$-value combination methods. The Bonferroni, Simes, and ACAT methods combine the marginal $p$-values from the eight molars, whereas WV-M-IC1 and WV-M-IC2 jointly analyze the eight molar-specific interval-censored outcomes. Differences in the resulting rankings reflect the distinct ways in which the two approaches use information across the eight molars. The rankings were also compared across the proportional hazards ($r=0$) and proportional odds ($r=1$) models to assess the robustness of the results to the choice of transformation model.

To further investigate the functional relevance of the regions prioritized by the association analyses, we conducted over-representation analysis (ORA) \citep{boyle2004go}, gene set enrichment analysis (GSEA) \citep{subramanian2005gene}, and network-based set enrichment analysis (NSEA) using the R package \texttt{clusterProfiler} \citep{xu2024using}. For these analyses, genes were ranked or selected according to their association results from the corresponding gene-based analyses. The conventional WV-IC-based approaches (Bonferroni, Simes, and ACAT) and the proposed WV-M-IC1 and WV-M-IC2 methods were evaluated separately to assess whether the multivariate analyses identified biologically coherent signals that were not apparent from marginal $p$-value combination approaches. The gene lists derived from the conventional combination approaches did not yield statistically significant pathways after multiple-testing adjustment. In contrast, the multivariate analyses identified several enriched biological processes and cellular components. Under WV-M-IC1, the prioritized genes were enriched for post-translational protein modification processes, including peptidyl-amino acid modification, peptidyl-lysine modification, and peptidyl-lysine acetylation, as well as cellular components including the early endosome and transferase complexes. Under WV-M-IC2, enriched pathways included structural cellular components, most notably keratin filament pathways. These findings suggest that the proposed multivariate association tests may capture biologically coherent genetic signals across the multiple primary molars that are less apparent when the molar-specific association results are analyzed through conventional $p$-value combination approaches.

\section{Discussion}
Large-scale genetic data have provided new opportunities to study the genetic basis of complex diseases. However, many biomedical outcomes are interval-censored because disease status is only observed at periodic visits. In this setting, the exact time of disease onset is unknown and can only be determined to fall between two observation times. Treating interval-censored outcomes as other types of survival data can lead to loss of information and reduced statistical efficiency. When multiple related phenotypes are measured for the same individual, analyzing each phenotype separately also does not fully use the information shared across outcomes. These issues motivated the development of the proposed methods.

In this study, we developed two Weighted V Test procedures for Multivariate Interval-Censored Data (WV-M-IC). The proposed methods extend the weighted V framework to multiple correlated interval-censored survival phenotypes. By analyzing multiple outcomes jointly, the proposed methods can use information shared across outcomes and can have higher power than analyzing each phenotype separately. The two proposed procedures provide different ways to construct the test statistic and its null distribution while retaining the main advantages of the weighted V framework.

In the simulation studies, we evaluated the type I error and power of the proposed methods under different outcome correlation structures and genetic effect settings. The proposed methods maintained appropriate type I error control across the scenarios considered. In terms of power, jointly analyzing multiple outcomes generally provided an advantage over analyzing each outcome separately when the outcomes shared genetic information. The amount of power gain depended on the correlation among outcomes and the pattern of genetic effects. These results show that information from correlated interval-censored outcomes can be useful for genetic association testing.

We applied the proposed methods to the ZOE 2.0 dataset for a gene-based analysis of ECC progression. No individual gene reached genome-wide significance after multiple-testing adjustment. However, several genes showed similar association patterns across multiple tooth-specific phenotypes. In particular, the WVMIC 2 test identified \textit{MAN1A2}, \textit{KRTAP3-3}, and \textit{KRTAP3-2} as showing consistent association patterns with age at ECC onset across molars. These genes were also included in several pathways identified by the downstream enrichment analysis, including post-translational modification and keratin filament organization. These results provide some biological support for the observed genetic signals. However, because the individual gene associations did not reach genome-wide significance, these findings should be considered exploratory. Larger studies and independent datasets are needed to further evaluate these candidate genes and pathways.

There are several limitations of the proposed methods. First, computational efficiency can become an issue when the number of individuals or genetic variants is large. This may be especially important when the proposed tests are applied to a large number of genes or variant sets in a genome-wide analysis. More efficient implementations, such as implementations using faster programming languages or optimized algorithms, could reduce the computational burden. This is an important direction for future work.

Another possible extension is to incorporate genetic heterogeneity into the multivariate interval-censored framework. The effects of a marker set may differ across subpopulations or individuals with different genetic backgrounds. The weighted V framework has previously been extended to account for such heterogeneity by incorporating a measure of similarity between individuals into the kernel function. A similar heterogeneity-weighted approach could be considered for WV-M-IC by modifying the genetic similarity measure to account for observed or latent population structure. Such an extension would allow the proposed multivariate framework to be used when genetic effects are not homogeneous across individuals or subpopulations.

Overall, the proposed WV-M-IC procedures extend the weighted V framework to multiple correlated interval-censored survival phenotypes and allow information from multiple outcomes to be used jointly. The simulation studies and the application to ECC show the potential usefulness of the proposed framework for genetic association analysis. Future work could extend the methods to account for genetic effect heterogeneity, improve computational efficiency, and evaluate their performance in larger and independent datasets.

\section*{Funding}
This work was supported in part by the National Institutes of Health (R03DE032357 to J.L., K.X., J.Z., Q.L. and C.L.).

\section*{Acknowledgments}
The data for the application presented in this work were from the ZOE 2.0 study, "Genome-wide association study of early childhood caries", which was supported by a grant
from the National Institute of Dental \& Craniofacial Research: U01-DE025046. Genotyping was performed by
CIDR, supported by a resource-allocation grant X01-HG010871, funded by NIDCR. We are grateful to the Principal Investigator, Kimon Divaris, for facilitating the data transfer and for helpful discussions on data aspects.

\section*{Data availability}
The data that support the findings of this study are available in dbGaP
(Accession No.: phs002232.v1.p1, “TOPDECC-Trans-omics for Precision 
Dentistry and Early Childhood Caries: Genome-Wide Genotyping 
(CIDR) and Microbiome in the ZOE 2.0 Study”).

\newpage
\appendix
\section{Power formulas for univariate and multivariate weighted V tests in linear models}

We consider a multivariate statistical framework to analyze the relationship between the response variables \( Y_1 \) and \( Y_2 \) and the genetic marker \( G \), accounting for covariates \( Z \). The model is specified as:
\[
\begin{aligned}
Y_1 &= \mu_1 + \alpha_1^T Z + \beta_1 G + \epsilon_1, \\
Y_2 &= \mu_2 + \alpha_2^T Z + \beta_2 G + \epsilon_2,
\end{aligned}
\]
where
\[
\begin{small}
\begin{bmatrix}
\epsilon_1 \\ \epsilon_2
\end{bmatrix}
\Big| Z, G \sim N \left(
\begin{bmatrix}
0 \\ 0
\end{bmatrix},
\begin{bmatrix}
\sigma_1^2 & \rho \sigma_1 \sigma_2 \\
\rho \sigma_1 \sigma_2 & \sigma_2^2
\end{bmatrix}
\right)
\end{small}
\]
with \( Z \perp G \) and \( E(Z) = 0 \).
\subsection{WV-IC Test}

To investigate the genetic effect of \( G \) on \( Y_1 \) and \( Y_2 \), we consider the hypotheses
\[
H_0: \beta_1 = 0 \quad \text{vs} \quad H_1: \beta_1 = \frac{1}{\sqrt{n}} h_1,
\]
where \( h_1 \) is fixed constants and \( n \) is the sample size. The test statistic is defined as
\[
n V^* = W_{\text{uni}}^T W_{\text{uni}},
\]
where
\(\begin{small}
W_{\text{uni}} = \frac{1}{\sqrt{n}} \sum_{i=1}^n 
\frac{Y_{i1} - \gamma_{10} - \gamma_1^T Z_i}{\sigma_1^2}
(G_i - E(G)), \quad 
\end{small}
(\gamma_{10},\gamma_{1}) = \arg\min_{\gamma_0, \gamma_1} E(Y_1 - \gamma_0 - \gamma^T Z)^2 \text{ and } \sigma_1^2=E(Y_1-\gamma_{10}-\gamma_1^T Z)^2. \) It can be shown \(\gamma_{10} = \mu_1 + \beta_1 E(G)\), \(\gamma_1 = \alpha_1\) and \(\sigma_1^2=\beta_1^2\sigma_a^2+ \sigma_1^2. \)
Under \( H_0 \), \( W_{\text{uni}} \xrightarrow{d} N \left( 0, \frac{\sigma_a^2}{\sigma_1^2} \right) \), while under \( H_1 \), \(W_{\text{multi}_1} \xrightarrow{d} N \left( \frac{h_1\sigma_a^2}{\sigma_1^2}, \frac{\sigma_a^2}{\sigma_1^2} \right) \).

The power of the univariate test is
\[
\text{Power}(\text{uni}) = P \left( 
 \frac{h_1\sigma_a^2}{\sigma_1^2} \chi_{1}^{2} 
> \chi_{1,\alpha}^2 \right).
\]
\subsection{WV-M-IC Test 1}

To investigate the genetic effect of \( G \) on \( Y_1 \) and \( Y_2 \), we consider the hypotheses
\[
H_0: \beta_1 = \beta_2 = 0 \quad \text{vs} \quad H_1: \beta_1 = \frac{1}{\sqrt{n}} h_1,\quad \beta_2 = \frac{1}{\sqrt{n}} h_2,\quad \max(h_1, h_2) > 0,
\]
where \( h_1, h_2 \) are fixed constants and \( n \) is the sample size. The test statistic is defined as
\[
n V^* = W_{\text{multi}_1}^T W_{\text{multi}_1},
\]
where
\[
\begin{small}
W_{\text{multi}_1} = \frac{1}{\sqrt{n}} \sum_{i=1}^n 
\begin{bmatrix}
\frac{Y_{i1} - \gamma_{10} - \gamma_1^T Z_i}{\sigma_1^2} \\
\frac{Y_{i2} - \gamma_{20} - \gamma_2^T Z_i}{\sigma_2^2}
\end{bmatrix}
(G_i - E(G)).
\end{small}
\]

Under \( H_0 \), \( W_{\text{multi}_1} \xrightarrow{d} N \left( 
\begin{bmatrix}
0 \\ 0
\end{bmatrix}, 
\Sigma_S \otimes \sigma_a^2 
\right) \), while under \( H_1 \), 
\(W_{\text{multi}_1} \xrightarrow{d} N \left( 
\begin{bmatrix}
\frac{h_1}{\sigma_1^2} \\
\frac{h_2}{\sigma_2^2}
\end{bmatrix} \sigma_a^2, 
\Sigma_S \otimes \sigma_a^2 
\right).\)

The eigen-decomposition of the covariance matrix is
\[
\Sigma_S = U \Lambda U^T = U 
\begin{bmatrix}
\lambda_1 & 0 \\
0 & \lambda_2
\end{bmatrix} 
U^T.
\]

We define
\[
\begin{bmatrix}
\delta_1 \\
\delta_2
\end{bmatrix}
= \Lambda^{-1/2} U^T 
\begin{bmatrix}
\frac{h_1}{\sigma_1^2} \\
\frac{h_2}{\sigma_2^2}
\end{bmatrix}
\sigma_a.
\]

The power of the test is
\[
\text{Power}(\text{Multi}_1) = P \left( 
\lambda_1 \chi_{1, \delta_1^2}^{(1)^2} + 
\lambda_2 \chi_{1, \delta_2^2}^{(2)^2} 
> F^{-1}_{\lambda_1 \chi_{1}^{(1)^2} + 
\lambda_2 \chi_{1}^{(2)^2}} (\alpha) \right),
\]
where \( \chi_{1, \delta_k^2}^{(k)^2} \) is a non-central chi-squared variable with 1 d.f. and non-centrality parameter \( \delta_k^2 \), and \( F^{-1} \) denotes the inverse CDF of the null distribution.
\subsection{WV-M-IC Test 2}

To investigate the genetic effect of \( G \) on \( Y_1 \) and \( Y_2 \), we consider the hypotheses
\[
H_0: \beta_1 = \beta_2 = 0 \quad \text{vs} \quad H_1: \beta_1 = \frac{1}{\sqrt{n}} h_1,\quad \beta_2 = \frac{1}{\sqrt{n}} h_2,\quad \max(h_1, h_2) > 0,
\]
where \( h_1, h_2 \) are fixed constants and \( n \) is the sample size. The test statistic is defined as
\[
n V^* = W_{\text{multi}_2}^T W_{\text{multi}_2},
\]
where
\[
\begin{small}
W_{\text{multi}_2} = \frac{1}{\sqrt{n}} \sum_{i=1}^n 
\begin{bmatrix}
\sigma_1 & 0 \\
0 & \sigma_2
\end{bmatrix}
\begin{bmatrix}
\frac{Y_{i1} - \gamma_{10} - \gamma_1^T Z_i}{\sigma_1^2} \\
\frac{Y_{i2} - \gamma_{20} - \gamma_2^T Z_i}{\sigma_2^2}
\end{bmatrix}
(G_i - E(G)).
\end{small}
\]

Under \( H_0 \), \( W_{\text{multi}_2} \xrightarrow{d} N \left( 
\begin{bmatrix}
0 \\ 0
\end{bmatrix}, 
\begin{bmatrix}
\frac{\sigma_{(1)}^2}{\sigma_1^2} & \frac{\rho\sigma_{(1)}\sigma_{(2)}}{\sigma_1\sigma_1}\\
\frac{\rho\sigma_{(1)}\sigma_{(2)}}{\sigma_1\sigma_1} & \frac{\sigma_{(2)}^2}{\sigma_2^2}
\end{bmatrix} 
\right) 
= N \left( 
\begin{bmatrix}
0 \\ 0
\end{bmatrix}, 
\begin{bmatrix}
1 & \rho \\
\rho & 1
\end{bmatrix} \sigma_a^2
\right) 
\), while under \( H_1 \) 
\(W_{\text{multi}_2} \xrightarrow{d} N \left( 
\begin{bmatrix}
\frac{h_1}{\sigma_1^2} \\
\frac{h_2}{\sigma_2^2}
\end{bmatrix} \sigma_a^2, 
\begin{bmatrix}
1 & \rho \\
\rho & 1
\end{bmatrix} \sigma_a^2
\right).\)

The power of the test is
\[
\text{Power}(\text{Multi}_2) = P \left( 
\tilde\lambda_1 \chi_{1, \tilde\delta_1^2}^{(1)^2} + 
\tilde\lambda_2 \chi_{1, \tilde\delta_2^2}^{(2)^2} 
> F^{-1}_{\tilde\lambda_1 \chi_{1}^{(1)^2} + 
\tilde\lambda_2 \chi_{1}^{(2)^2}} (\alpha) \right),
\]
where \(\begin{bmatrix}
\tilde\delta_1 \\
\tilde\delta_2
\end{bmatrix}
= \tilde\Lambda^{-1/2} \tilde{U}^T 
\begin{bmatrix}
\frac{h_1}{\sigma_1^2} \\
\frac{h_2}{\sigma_2^2}
\end{bmatrix}
\sigma_a.\)
and 
\(\begin{bmatrix}
1 & \rho \\
\rho & 1
\end{bmatrix}
=
\tilde{U} \tilde\Lambda U^T = \tilde{U} 
\begin{bmatrix}
\tilde\lambda_1 & 0 \\
0 & \tilde\lambda_2
\end{bmatrix} 
\tilde{U}^T.
\)

\newpage
\bibliography{WV-M-IC_ref}

@article{li2021set,
  title={Set-based genetic association and interaction tests for survival outcomes based on weighted V statistics},
  author={Li, Chenxi and Wu, Di and Lu, Qing},
  journal={Genetic epidemiology},
  volume={45},
  number={1},
  pages={46--63},
  year={2021},
  publisher={Wiley Online Library}
}

@article{wu2021multi,
  title={Multi-marker genetic association and interaction tests with interval-censored survival outcomes},
  author={Wu, Di and Li, Chenxi and Lu, Qing},
  journal={Genetic epidemiology},
  volume={45},
  number={8},
  pages={860--873},
  year={2021},
  publisher={Wiley Online Library}
}

@article{choi2024variance,
  title={Variance-components tests for genetic association with multiple interval-censored outcomes},
  author={Choi, Jaihee and Xu, Zhichao and Sun, Ryan},
  journal={Statistics in Medicine},
  volume={43},
  number={13},
  pages={2560--2574},
  year={2024},
  publisher={Wiley Online Library}
}

@article{davies1980algorithm,
  title={Algorithm AS 155: The distribution of a linear combination of $\chi$ 2 random variables},
  author={Davies, Robert B},
  journal={Applied Statistics},
  pages={323--333},
  year={1980},
  publisher={JSTOR}
}

@article{ginnis2019measurement,
  title={Measurement of early childhood oral health for research purposes: dental caries experience and developmental defects of the enamel in the primary dentition},
  author={Ginnis, Jeannie and Ferreira Zandon{\'a}, Andrea G and Slade, Gary D and Cantrell, John and Antonio, Mikafui E and Pahel, Bhavna T and Meyer, Beau D and Shrestha, Poojan and Simancas-Pallares, Miguel A and Joshi, Ashwini R and others},
  journal={Odontogenesis: Methods and Protocols},
  pages={511--523},
  year={2019},
  publisher={Springer}
}

@article{divaris2020cohort,
  title={Cohort profile: ZOE 2.0—a community-based genetic epidemiologic study of early childhood oral health},
  author={Divaris, Kimon and Slade, Gary D and Ferreira Zandona, Andrea G and Preisser, John S and Ginnis, Jeannie and Simancas-Pallares, Miguel A and Agler, Cary S and Shrestha, Poojan and Karhade, Deepti S and Ribeiro, Apoena de Aguiar and others},
  journal={International journal of environmental research and public health},
  volume={17},
  number={21},
  pages={8056},
  year={2020},
  publisher={MDPI}
}

@article{bland1995multiple,
  title={Multiple significance tests: the Bonferroni method},
  author={Bland, J Martin and Altman, Douglas G},
  journal={Bmj},
  volume={310},
  number={6973},
  pages={170},
  year={1995},
  publisher={British Medical Journal Publishing Group}
}

@article{simes1986improved,
  title={An improved Bonferroni procedure for multiple tests of significance},
  author={Simes, R John},
  journal={Biometrika},
  volume={73},
  number={3},
  pages={751--754},
  year={1986},
  publisher={Oxford University Press}
}

@article{liu2019acat,
  title={ACAT: a fast and powerful p value combination method for rare-variant analysis in sequencing studies},
  author={Liu, Yaowu and Chen, Sixing and Li, Zilin and Morrison, Alanna C and Boerwinkle, Eric and Lin, Xihong},
  journal={The American Journal of Human Genetics},
  volume={104},
  number={3},
  pages={410--421},
  year={2019},
  publisher={Elsevier}
}

@article{benjamini1995controlling,
  title={Controlling the false discovery rate: a practical and powerful approach to multiple testing},
  author={Benjamini, Yoav and Hochberg, Yosef},
  journal={Journal of the Royal statistical society: series B (Methodological)},
  volume={57},
  number={1},
  pages={289--300},
  year={1995},
  publisher={Wiley Online Library}
}

@article{zeng2016maximum,
  title={Maximum likelihood estimation for semiparametric transformation models with interval-censored data},
  author={Zeng, Donglin and Mao, Lu and Lin, DY},
  journal={Biometrika},
  volume={103},
  number={2},
  pages={253--271},
  year={2016},
  publisher={Oxford University Press}
}

@article{wei2017generalized,
  title={A generalized association test based on U statistics},
  author={Wei, Changshuai and Lu, Qing},
  journal={Bioinformatics},
  volume={33},
  number={13},
  pages={1963--1971},
  year={2017},
  publisher={Oxford University Press}
}

@article{xu2024using,
  title={Using clusterProfiler to characterize multiomics data},
  author={Xu, Shuangbin and Hu, Erqiang and Cai, Yantong and Xie, Zijing and Luo, Xiao and Zhan, Li and Tang, Wenli and Wang, Qianwen and Liu, Bingdong and Wang, Rui and others},
  journal={Nature protocols},
  volume={19},
  number={11},
  pages={3292--3320},
  year={2024},
  publisher={Nature Publishing Group UK London}
}

@article{boyle2004go,
  title={GO:: TermFinder—open source software for accessing Gene Ontology information and finding significantly enriched Gene Ontology terms associated with a list of genes},
  author={Boyle, Elizabeth I and Weng, Shuai and Gollub, Jeremy and Jin, Heng and Botstein, David and Cherry, J Michael and Sherlock, Gavin},
  journal={Bioinformatics},
  volume={20},
  number={18},
  pages={3710--3715},
  year={2004},
  publisher={Oxford University Press}
}

@article{subramanian2005gene,
  title={Gene set enrichment analysis: a knowledge-based approach for interpreting genome-wide expression profiles},
  author={Subramanian, Aravind and Tamayo, Pablo and Mootha, Vamsi K and Mukherjee, Sayan and Ebert, Benjamin L and Gillette, Michael A and Paulovich, Amanda and Pomeroy, Scott L and Golub, Todd R and Lander, Eric S and others},
  journal={Proceedings of the national academy of sciences},
  volume={102},
  number={43},
  pages={15545--15550},
  year={2005},
  publisher={National Academy of Sciences}
}

@article{liu2008estimation,
  title={Estimation and testing for the effect of a genetic pathway on a disease outcome using logistic kernel machine regression via logistic mixed models},
  author={Liu, Dawei and Ghosh, Debashis and Lin, Xihong},
  journal={BMC bioinformatics},
  volume={9},
  number={1},
  pages={292},
  year={2008},
  publisher={Springer}
}

@article{liu2007semiparametric,
  title={Semiparametric regression of multidimensional genetic pathway data: Least-squares kernel machines and linear mixed models},
  author={Liu, Dawei and Lin, Xihong and Ghosh, Debashis},
  journal={Biometrics},
  volume={63},
  number={4},
  pages={1079--1088},
  year={2007},
  publisher={Wiley Online Library}
}

@article{wu2011rare,
  title={Rare-variant association testing for sequencing data with the sequence kernel association test},
  author={Wu, Michael C and Lee, Seunggeun and Cai, Tianxi and Li, Yun and Boehnke, Michael and Lin, Xihong},
  journal={The American Journal of Human Genetics},
  volume={89},
  number={1},
  pages={82--93},
  year={2011},
  publisher={Elsevier}
}

@article{li2022multi,
  title={A multi-dimensional integrative scoring framework for predicting functional variants in the human genome},
  author={Li, Xihao and Yung, Godwin and Zhou, Hufeng and Sun, Ryan and Li, Zilin and Hou, Kangcheng and Zhang, Martin Jinye and Liu, Yaowu and Arapoglou, Theodore and Wang, Chen and others},
  journal={The American Journal of Human Genetics},
  volume={109},
  number={3},
  pages={446--456},
  year={2022},
  publisher={Elsevier}
}

\newpage 
\begin{table}[htbp]
\centering
\small
\begin{threeparttable}
\caption{Empirical size and power of univariate and multivariate tests (n=400, p=15). Survival times are generated under the proportional hazards model (r=0).}
\begin{tabular}{llcccccccc}
\toprule
\multirow{3}{*}{} & \multirow{3}{*}{} 
& \multicolumn{2}{c}{\( \rho \) = 0.1} 
& \multicolumn{2}{c}{\( \rho \) = 0.25} 
& \multicolumn{2}{c}{\( \rho \) = 0.5} 
& \multicolumn{2}{c}{\( \rho \) = 0.75} \\
\cmidrule(lr){3-4} \cmidrule(lr){5-6} \cmidrule(lr){7-8} \cmidrule(lr){9-10}
& & Size & Power & Size & Power & Size & Power & Size & Power \\
\midrule
\multicolumn{10}{l}{\textbf{Univariate Tests}} \\
& WV-IC test for T1 & 0.051 & 0.301 & 0.055 & 0.297 & 0.044 & 0.292 & 0.042 & 0.288 \\
& WV-IC test for T2 & 0.056 & 0.287 & 0.051 & 0.302 & 0.050 & 0.294 & 0.040 & 0.297 \\
& WV-IC test for T3 & 0.049 & 0.173 & 0.052 & 0.179 & 0.045 & 0.172 & 0.045 & 0.187 \\
& WV-IC test for T4 & 0.044 & 0.174 & 0.042 & 0.184 & 0.042 & 0.175 & 0.048 & 0.178 \\
& WV-IC test for T5 & 0.053 & 0.118 & 0.055 & 0.110 & 0.051 & 0.124 & 0.043 & 0.116 \\
& WV-IC test for T6 & 0.048 & 0.108 & 0.054 & 0.110 & 0.057 & 0.119 & 0.053 & 0.113 \\
& WV-IC test for T7 & 0.062 & 0.073 & 0.058 & 0.066 & 0.054 & 0.067 & 0.052 & 0.064 \\
& WV-IC test for T8 & 0.057 & 0.069 & 0.047 & 0.056 & 0.050 & 0.056 & 0.049 & 0.063 \\
\midrule
\multicolumn{10}{l}{\textbf{Multivariate Tests (Scenario 1)}} \\
& WV-M-IC1     & 0.046 & 0.386 & 0.051 & 0.313 & 0.045 & 0.240 & 0.038 & 0.216 \\
& WV-M-IC2     & 0.048 & 0.432 & 0.043 & 0.411 & 0.044 & 0.358 & 0.048 & 0.317 \\
& Bonferroni correction & 0.051 & 0.323 & 0.046 & 0.317 & 0.044 & 0.290 & 0.045 & 0.268 \\
& Simes test      & 0.051 & 0.338 & 0.046 & 0.333 & 0.045 & 0.305 & 0.045 & 0.282 \\
& ACAT                 & 0.044 & 0.375 & 0.049 & 0.351 & 0.047 & 0.333 & 0.055 & 0.317 \\
\midrule
\multicolumn{10}{l}{\textbf{Multivariate Tests (Scenario 2)}} \\
& WV-M-IC1     & 0.057 & 0.314 & 0.053 & 0.248 & 0.048 & 0.235 & 0.043 & 0.260 \\
& WV-M-IC2     & 0.054 & 0.406 & 0.056 & 0.358 & 0.046 & 0.293 & 0.050 & 0.234 \\
& Bonferroni correction & 0.050 & 0.285 & 0.057 & 0.263 & 0.044 & 0.242 & 0.036 & 0.229 \\
& Simes test      & 0.050 & 0.293 & 0.058 & 0.272 & 0.045 & 0.254 & 0.038 & 0.237 \\
& ACAT                 & 0.054 & 0.315 & 0.058 & 0.304 & 0.053 & 0.285 & 0.062 & 0.277 \\
\bottomrule
\end{tabular}
\end{threeparttable}
\end{table}

\newpage 
\begin{table}[htbp]
\centering
\small
\begin{threeparttable}
\caption{Empirical size and power of univariate and multivariate tests (n=400, p=25). Survival times are generated under the proportional hazards model (r=0).}
\begin{tabular}{llcccccccc}
\toprule
\multirow{3}{*}{} & \multirow{3}{*}{} 
& \multicolumn{2}{c}{\( \rho \) = 0.1} 
& \multicolumn{2}{c}{\( \rho \) = 0.25} 
& \multicolumn{2}{c}{\( \rho \) = 0.5} 
& \multicolumn{2}{c}{\( \rho \) = 0.75} \\
\cmidrule(lr){3-4} \cmidrule(lr){5-6} \cmidrule(lr){7-8} \cmidrule(lr){9-10}
& & Size & Power & Size & Power & Size & Power & Size & Power \\
\midrule
\multicolumn{10}{l}{\textbf{Univariate Tests}} \\
& WV-IC test for T1 & 0.050 & 0.485 & 0.051 & 0.499 & 0.045 & 0.500 & 0.042 & 0.494 \\
& WV-IC test for T2 & 0.044 & 0.496 & 0.053 & 0.494 & 0.044 & 0.505 & 0.047 & 0.490 \\
& WV-IC test for T3 & 0.055 & 0.325 & 0.058 & 0.316 & 0.052 & 0.313 & 0.055 & 0.332 \\
& WV-IC test for T4 & 0.038 & 0.314 & 0.039 & 0.313 & 0.035 & 0.319 & 0.051 & 0.309 \\
& WV-IC test for T5 & 0.056 & 0.178 & 0.049 & 0.192 & 0.045 & 0.176 & 0.044 & 0.167 \\
& WV-IC test for T6 & 0.045 & 0.169 & 0.051 & 0.169 & 0.060 & 0.177 & 0.054 & 0.178 \\
& WV-IC test for T7 & 0.060 & 0.083 & 0.057 & 0.085 & 0.057 & 0.085 & 0.053 & 0.073 \\
& WV-IC test for T8 & 0.054 & 0.079 & 0.054 & 0.076 & 0.054 & 0.066 & 0.049 & 0.072 \\
\midrule
\multicolumn{10}{l}{\textbf{Multivariate Tests (Scenario 1)}} \\
& WV-M-IC1     & 0.044 & 0.584 & 0.042 & 0.517 & 0.047 & 0.455 & 0.044 & 0.405 \\
& WV-M-IC2     & 0.046 & 0.642 & 0.042 & 0.629 & 0.041 & 0.582 & 0.047 & 0.524 \\
& Bonferroni correction & 0.047 & 0.547 & 0.048 & 0.527 & 0.046 & 0.493 & 0.044 & 0.467 \\
& Simes test      & 0.047 & 0.560 & 0.050 & 0.537 & 0.046 & 0.504 & 0.045 & 0.481 \\
& ACAT                 & 0.045 & 0.582 & 0.048 & 0.566 & 0.050 & 0.542 & 0.054 & 0.514 \\
\midrule
\multicolumn{10}{l}{\textbf{Multivariate Tests (Scenario 2)}} \\
& WV-M-IC1     & 0.057 & 0.542 & 0.052 & 0.479 & 0.057 & 0.432 & 0.050 & 0.487 \\
& WV-M-IC2     & 0.060 & 0.618 & 0.055 & 0.583 & 0.050 & 0.500 & 0.049 & 0.411 \\
& Bonferroni correction & 0.057 & 0.496 & 0.051 & 0.475 & 0.046 & 0.443 & 0.038 & 0.404 \\
& Simes test      & 0.059 & 0.515 & 0.052 & 0.492 & 0.047 & 0.448 & 0.045 & 0.419 \\
& ACAT                 & 0.060 & 0.541 & 0.065 & 0.530 & 0.053 & 0.481 & 0.063 & 0.462 \\
\bottomrule
\end{tabular}
\end{threeparttable}
\end{table}

\newpage 
\begin{table}[htbp]
\centering
\small
\begin{threeparttable}
\caption{Empirical size and power of univariate and multivariate tests (n=800, p=15). Survival times are generated under the proportional hazards model (r=0).}
\begin{tabular}{llcccccccc}
\toprule
\multirow{3}{*}{} & \multirow{3}{*}{} 
& \multicolumn{2}{c}{\( \rho \) = 0.1} 
& \multicolumn{2}{c}{\( \rho \) = 0.25} 
& \multicolumn{2}{c}{\( \rho \) = 0.5} 
& \multicolumn{2}{c}{\( \rho \) = 0.75} \\
\cmidrule(lr){3-4} \cmidrule(lr){5-6} \cmidrule(lr){7-8} \cmidrule(lr){9-10}
& & Size & Power & Size & Power & Size & Power & Size & Power \\
\midrule
\multicolumn{10}{l}{\textbf{Univariate Tests}} \\
& WV-IC test for T1 & 0.054 & 0.472 & 0.048 & 0.479 & 0.050 & 0.486 & 0.045 & 0.476 \\
& WV-IC test for T2 & 0.045 & 0.489 & 0.053 & 0.477 & 0.046 & 0.463 & 0.050 & 0.471 \\
& WV-IC test for T3 & 0.057 & 0.321 & 0.058 & 0.328 & 0.049 & 0.321 & 0.051 & 0.315 \\
& WV-IC test for T4 & 0.055 & 0.299 & 0.055 & 0.297 & 0.049 & 0.303 & 0.047 & 0.301 \\
& WV-IC test for T5 & 0.041 & 0.172 & 0.032 & 0.165 & 0.043 & 0.169 & 0.037 & 0.160 \\
& WV-IC test for T6 & 0.058 & 0.178 & 0.053 & 0.169 & 0.066 & 0.172 & 0.055 & 0.184 \\
& WV-IC test for T7 & 0.053 & 0.071 & 0.049 & 0.077 & 0.050 & 0.083 & 0.044 & 0.074 \\
& WV-IC test for T8 & 0.034 & 0.068 & 0.040 & 0.069 & 0.045 & 0.072 & 0.036 & 0.073 \\
\midrule
\multicolumn{10}{l}{\textbf{Multivariate Tests (Scenario 1)}} \\
& WV-M-IC1     & 0.044 & 0.603 & 0.051 & 0.530 & 0.055 & 0.438 & 0.049 & 0.403 \\
& WV-M-IC2     & 0.046 & 0.640 & 0.045 & 0.624 & 0.048 & 0.577 & 0.044 & 0.504 \\
& Bonferroni correction & 0.059 & 0.544 & 0.045 & 0.531 & 0.051 & 0.497 & 0.035 & 0.461 \\
& Simes test      & 0.059 & 0.565 & 0.045 & 0.544 & 0.052 & 0.509 & 0.038 & 0.476 \\
& ACAT                 & 0.062 & 0.595 & 0.051 & 0.570 & 0.055 & 0.545 & 0.046 & 0.509 \\
\midrule
\multicolumn{10}{l}{\textbf{Multivariate Tests (Scenario 2)}} \\
& WV-M-IC1     & 0.047 & 0.528 & 0.040 & 0.444 & 0.050 & 0.412 & 0.045 & 0.462 \\
& WV-M-IC2     & 0.051 & 0.615 & 0.043 & 0.566 & 0.048 & 0.479 & 0.049 & 0.386 \\
& Bonferroni correction & 0.058 & 0.479 & 0.048 & 0.462 & 0.041 & 0.434 & 0.032 & 0.398 \\
& Simes test      & 0.058 & 0.493 & 0.048 & 0.482 & 0.047 & 0.451 & 0.036 & 0.414 \\
& ACAT                 & 0.053 & 0.522 & 0.048 & 0.515 & 0.054 & 0.484 & 0.050 & 0.441 \\
\bottomrule
\end{tabular}
\end{threeparttable}
\end{table}

\newpage 
\begin{table}[htbp]
\centering
\small
\begin{threeparttable}
\caption{Empirical size and power of univariate and multivariate tests (n=800, p=25). Survival times are generated under the proportional hazards model (r=0).}
\begin{tabular}{llcccccccc}
\toprule
\multirow{3}{*}{} & \multirow{3}{*}{} 
& \multicolumn{2}{c}{\( \rho \) = 0.1} 
& \multicolumn{2}{c}{\( \rho \) = 0.25} 
& \multicolumn{2}{c}{\( \rho \) = 0.5} 
& \multicolumn{2}{c}{\( \rho \) = 0.75} \\
\cmidrule(lr){3-4} \cmidrule(lr){5-6} \cmidrule(lr){7-8} \cmidrule(lr){9-10}
& & Size & Power & Size & Power & Size & Power & Size & Power \\
\midrule
\multicolumn{10}{l}{\textbf{Univariate Tests}} \\
& WV-IC test for T1 & 0.058 & 0.697 & 0.049 & 0.692 & 0.053 & 0.708 & 0.046 & 0.699 \\
& WV-IC test for T2 & 0.039 & 0.687 & 0.048 & 0.691 & 0.043 & 0.690 & 0.053 & 0.676 \\
& WV-IC test for T3 & 0.050 & 0.526 & 0.062 & 0.528 & 0.058 & 0.533 & 0.045 & 0.515 \\
& WV-IC test for T4 & 0.053 & 0.504 & 0.056 & 0.509 & 0.049 & 0.519 & 0.048 & 0.517 \\
& WV-IC test for T5 & 0.043 & 0.298 & 0.033 & 0.291 & 0.044 & 0.280 & 0.040 & 0.294 \\
& WV-IC test for T6 & 0.053 & 0.318 & 0.050 & 0.306 & 0.053 & 0.305 & 0.055 & 0.296 \\
& WV-IC test for T7 & 0.050 & 0.102 & 0.051 & 0.103 & 0.043 & 0.103 & 0.046 & 0.099 \\
& WV-IC test for T8 & 0.049 & 0.110 & 0.052 & 0.108 & 0.049 & 0.118 & 0.034 & 0.103 \\
\midrule
\multicolumn{10}{l}{\textbf{Multivariate Tests (Scenario 1)}} \\
& WV-M-IC1     & 0.046 & 0.783 & 0.053 & 0.738 & 0.053 & 0.664 & 0.044 & 0.621 \\
& WV-M-IC2     & 0.046 & 0.822 & 0.047 & 0.802 & 0.044 & 0.781 & 0.043 & 0.724 \\
& Bonferroni correction & 0.063 & 0.741 & 0.046 & 0.742 & 0.052 & 0.712 & 0.033 & 0.679 \\
& Simes test      & 0.064 & 0.750 & 0.046 & 0.750 & 0.052 & 0.724 & 0.037 & 0.690 \\
& ACAT                 & 0.061 & 0.777 & 0.053 & 0.770 & 0.054 & 0.750 & 0.048 & 0.714 \\
\midrule
\multicolumn{10}{l}{\textbf{Multivariate Tests (Scenario 2)}} \\
& WV-M-IC1     & 0.049 & 0.725 & 0.052 & 0.665 & 0.044 & 0.633 & 0.044 & 0.675 \\
& WV-M-IC2     & 0.051 & 0.795 & 0.046 & 0.770 & 0.040 & 0.703 & 0.043 & 0.612 \\
& Bonferroni correction & 0.050 & 0.698 & 0.048 & 0.694 & 0.037 & 0.658 & 0.026 & 0.624 \\
& Simes test      & 0.053 & 0.711 & 0.052 & 0.710 & 0.040 & 0.671 & 0.029 & 0.637 \\
& ACAT                 & 0.052 & 0.734 & 0.051 & 0.737 & 0.051 & 0.702 & 0.048 & 0.664 \\
\bottomrule
\end{tabular}
\end{threeparttable}
\end{table}

\newpage 
\begin{table}[htbp]
\centering
\small
\begin{threeparttable}
\caption{Empirical size and power of univariate and multivariate tests (n=400, p=15). Survival times are generated from the proportional odds model (r=1).}
\begin{tabular}{llcccccccc}
\toprule
\multirow{3}{*}{} & \multirow{3}{*}{} 
& \multicolumn{2}{c}{\( \rho \) = 0.1} 
& \multicolumn{2}{c}{\( \rho \) = 0.25} 
& \multicolumn{2}{c}{\( \rho \) = 0.5} 
& \multicolumn{2}{c}{\( \rho \) = 0.75} \\
\cmidrule(lr){3-4} \cmidrule(lr){5-6} \cmidrule(lr){7-8} \cmidrule(lr){9-10}
& & Size & Power & Size & Power & Size & Power & Size & Power \\
\midrule
\multicolumn{10}{l}{\textbf{Univariate Tests}} \\
& WV-IC test for T1 & 0.048 & 0.278 & 0.046 & 0.279 & 0.045 & 0.278 & 0.042 & 0.283 \\
& WV-IC test for T2 & 0.052 & 0.276 & 0.052 & 0.293 & 0.052 & 0.288 & 0.047 & 0.287 \\
& WV-IC test for T3 & 0.051 & 0.171 & 0.050 & 0.168 & 0.045 & 0.153 & 0.048 & 0.180 \\
& WV-IC test for T4 & 0.049 & 0.170 & 0.045 & 0.173 & 0.040 & 0.163 & 0.048 & 0.164 \\
& WV-IC test for T5 & 0.049 & 0.116 & 0.058 & 0.111 & 0.053 & 0.124 & 0.048 & 0.114 \\
& WV-IC test for T6 & 0.044 & 0.113 & 0.053 & 0.112 & 0.061 & 0.116 & 0.051 & 0.110 \\
& WV-IC test for T7 & 0.058 & 0.071 & 0.057 & 0.066 & 0.048 & 0.058 & 0.045 & 0.069 \\
& WV-IC test for T8 & 0.052 & 0.070 & 0.058 & 0.066 & 0.044 & 0.062 & 0.051 & 0.059 \\
\midrule
\multicolumn{10}{l}{\textbf{Multivariate Tests (Scenario 1)}} \\
& WV-M-IC1 & 0.045 & 0.358 & 0.049 & 0.281 & 0.051 & 0.225 & 0.043 & 0.195 \\
& WV-M-IC2 & 0.043 & 0.409 & 0.046 & 0.397 & 0.042 & 0.338 & 0.050 & 0.294 \\
& Bonferroni correction & 0.047 & 0.313 & 0.048 & 0.285 & 0.043 & 0.274 & 0.046 & 0.253 \\
& Simes test & 0.049 & 0.324 & 0.050 & 0.297 & 0.044 & 0.282 & 0.047 & 0.265 \\
& ACAT & 0.046 & 0.355 & 0.057 & 0.327 & 0.045 & 0.311 & 0.051 & 0.308 \\
\midrule
\multicolumn{10}{l}{\textbf{Multivariate Tests (Scenario 2)}} \\
& WV-M-IC1 & 0.053 & 0.301 & 0.065 & 0.237 & 0.056 & 0.220 & 0.049 & 0.267 \\
& WV-M-IC2 & 0.054 & 0.391 & 0.052 & 0.343 & 0.045 & 0.274 & 0.050 & 0.208 \\
& Bonferroni correction & 0.051 & 0.271 & 0.052 & 0.252 & 0.042 & 0.223 & 0.038 & 0.201 \\
& Simes test & 0.051 & 0.284 & 0.053 & 0.267 & 0.046 & 0.234 & 0.041 & 0.217 \\
& ACAT & 0.052 & 0.314 & 0.055 & 0.295 & 0.051 & 0.272 & 0.065 & 0.264 \\
\bottomrule
\end{tabular}
\end{threeparttable}
\end{table}

\newpage 
\begin{table}[htbp]
\centering
\small
\begin{threeparttable}
\caption{Empirical size and power of univariate and multivariate tests (n=400, p=25). Survival times are generated from the proportional odds model (r=1).}
\begin{tabular}{llcccccccc}
\toprule
\multirow{3}{*}{} & \multirow{3}{*}{} 
& \multicolumn{2}{c}{\( \rho \) = 0.1} 
& \multicolumn{2}{c}{\( \rho \) = 0.25} 
& \multicolumn{2}{c}{\( \rho \) = 0.5} 
& \multicolumn{2}{c}{\( \rho \) = 0.75} \\
\cmidrule(lr){3-4} \cmidrule(lr){5-6} \cmidrule(lr){7-8} \cmidrule(lr){9-10}
& & Size & Power & Size & Power & Size & Power & Size & Power \\
\midrule
\multicolumn{10}{l}{\textbf{Univariate Tests}} \\
& WV-IC test for T1 & 0.046 & 0.465 & 0.047 & 0.460 & 0.052 & 0.454 & 0.043 & 0.460 \\
& WV-IC test for T2 & 0.048 & 0.458 & 0.054 & 0.461 & 0.049 & 0.476 & 0.051 & 0.464 \\
& WV-IC test for T3 & 0.055 & 0.318 & 0.055 & 0.312 & 0.047 & 0.291 & 0.052 & 0.313 \\
& WV-IC test for T4 & 0.048 & 0.291 & 0.044 & 0.292 & 0.045 & 0.285 & 0.052 & 0.302 \\
& WV-IC test for T5 & 0.047 & 0.173 & 0.053 & 0.186 & 0.051 & 0.177 & 0.049 & 0.164 \\
& WV-IC test for T6 & 0.051 & 0.166 & 0.047 & 0.165 & 0.056 & 0.170 & 0.057 & 0.178 \\
& WV-IC test for T7 & 0.065 & 0.089 & 0.058 & 0.093 & 0.050 & 0.081 & 0.041 & 0.074 \\
& WV-IC test for T8 & 0.051 & 0.077 & 0.053 & 0.072 & 0.048 & 0.075 & 0.049 & 0.076 \\
\midrule
\multicolumn{10}{l}{\textbf{Multivariate Tests (Scenario 1)}} \\
& WV-M-IC1 & 0.041 & 0.569 & 0.044 & 0.500 & 0.051 & 0.427 & 0.045 & 0.369 \\
& WV-M-IC2 & 0.044 & 0.612 & 0.048 & 0.596 & 0.046 & 0.550 & 0.053 & 0.487 \\
& Bonferroni correction & 0.051 & 0.524 & 0.052 & 0.502 & 0.049 & 0.471 & 0.046 & 0.431 \\
& Simes test & 0.051 & 0.536 & 0.053 & 0.510 & 0.049 & 0.482 & 0.048 & 0.444 \\
& ACAT & 0.044 & 0.572 & 0.053 & 0.550 & 0.052 & 0.520 & 0.059 & 0.491 \\
\midrule
\multicolumn{10}{l}{\textbf{Multivariate Tests (Scenario 2)}} \\
& WV-M-IC1 & 0.059 & 0.526 & 0.053 & 0.450 & 0.059 & 0.411 & 0.054 & 0.475 \\
& WV-M-IC2 & 0.062 & 0.596 & 0.054 & 0.545 & 0.050 & 0.466 & 0.052 & 0.359 \\
& Bonferroni correction & 0.050 & 0.478 & 0.056 & 0.455 & 0.041 & 0.411 & 0.039 & 0.373 \\
& Simes test & 0.052 & 0.495 & 0.057 & 0.467 & 0.045 & 0.424 & 0.044 & 0.384 \\
& ACAT & 0.056 & 0.514 & 0.062 & 0.498 & 0.050 & 0.462 & 0.062 & 0.423 \\
\bottomrule
\end{tabular}
\end{threeparttable}
\end{table}

\newpage 
\begin{table}[htbp]
\centering
\small
\begin{threeparttable}
\caption{Empirical size and power of univariate and multivariate tests (n=800, p=15). Survival times are generated from the proportional odds model (r=1).}
\begin{tabular}{llcccccccc}
\toprule
\multirow{3}{*}{} & \multirow{3}{*}{} 
& \multicolumn{2}{c}{\( \rho \) = 0.1} 
& \multicolumn{2}{c}{\( \rho \) = 0.25} 
& \multicolumn{2}{c}{\( \rho \) = 0.5} 
& \multicolumn{2}{c}{\( \rho \) = 0.75} \\
\cmidrule(lr){3-4} \cmidrule(lr){5-6} \cmidrule(lr){7-8} \cmidrule(lr){9-10}
& & Size & Power & Size & Power & Size & Power & Size & Power \\
\midrule
\multicolumn{10}{l}{\textbf{Univariate Tests}} \\
& WV-IC test for T1 & 0.044 & 0.467 & 0.040 & 0.463 & 0.040 & 0.459 & 0.041 & 0.443 \\
& WV-IC test for T2 & 0.055 & 0.455 & 0.051 & 0.447 & 0.066 & 0.444 & 0.058 & 0.452 \\
& WV-IC test for T3 & 0.048 & 0.300 & 0.052 & 0.297 & 0.046 & 0.291 & 0.047 & 0.299 \\
& WV-IC test for T4 & 0.062 & 0.314 & 0.045 & 0.296 & 0.042 & 0.279 & 0.045 & 0.290 \\
& WV-IC test for T5 & 0.049 & 0.158 & 0.053 & 0.155 & 0.055 & 0.152 & 0.056 & 0.161 \\
& WV-IC test for T6 & 0.045 & 0.157 & 0.045 & 0.166 & 0.043 & 0.159 & 0.045 & 0.149 \\
& WV-IC test for T7 & 0.034 & 0.068 & 0.030 & 0.063 & 0.035 & 0.059 & 0.051 & 0.068 \\
& WV-IC test for T8 & 0.055 & 0.092 & 0.056 & 0.082 & 0.053 & 0.077 & 0.071 & 0.079 \\
\midrule
\multicolumn{10}{l}{\textbf{Multivariate Tests (Scenario 1)}} \\
& WV-M-IC1 & 0.054 & 0.584 & 0.043 & 0.496 & 0.051 & 0.405 & 0.051 & 0.353 \\
& WV-M-IC2 & 0.053 & 0.643 & 0.040 & 0.616 & 0.039 & 0.537 & 0.037 & 0.469 \\
& Bonferroni correction & 0.044 & 0.539 & 0.047 & 0.513 & 0.045 & 0.466 & 0.035 & 0.406 \\
& Simes test & 0.047 & 0.551 & 0.048 & 0.529 & 0.047 & 0.479 & 0.039 & 0.426 \\
& ACAT & 0.055 & 0.589 & 0.047 & 0.555 & 0.048 & 0.511 & 0.047 & 0.469 \\
\midrule
\multicolumn{10}{l}{\textbf{Multivariate Tests (Scenario 2)}} \\
& WV-M-IC1 & 0.044 & 0.507 & 0.042 & 0.425 & 0.038 & 0.380 & 0.043 & 0.449 \\
& WV-M-IC2 & 0.045 & 0.594 & 0.040 & 0.546 & 0.047 & 0.431 & 0.049 & 0.331 \\
& Bonferroni correction & 0.050 & 0.481 & 0.043 & 0.448 & 0.033 & 0.412 & 0.033 & 0.345 \\
& Simes test & 0.050 & 0.496 & 0.044 & 0.459 & 0.033 & 0.422 & 0.037 & 0.359 \\
& ACAT & 0.047 & 0.531 & 0.047 & 0.490 & 0.048 & 0.463 & 0.056 & 0.401 \\
\bottomrule
\end{tabular}
\end{threeparttable}
\end{table}

\newpage 
\begin{table}[htbp]
\centering
\small
\begin{threeparttable}
\caption{Empirical size and power of univariate and multivariate tests (n=800, p=25). Survival times are generated from the proportional odds model (r=1).}
\begin{tabular}{llcccccccc}
\toprule
\multirow{3}{*}{} & \multirow{3}{*}{} 
& \multicolumn{2}{c}{\( \rho \) = 0.1} 
& \multicolumn{2}{c}{\( \rho \) = 0.25} 
& \multicolumn{2}{c}{\( \rho \) = 0.5} 
& \multicolumn{2}{c}{\( \rho \) = 0.75} \\
\cmidrule(lr){3-4} \cmidrule(lr){5-6} \cmidrule(lr){7-8} \cmidrule(lr){9-10}
& & Size & Power & Size & Power & Size & Power & Size & Power \\
\midrule
\multicolumn{10}{l}{\textbf{Univariate Tests}} \\
& WV-IC test for T1 & 0.045 & 0.677 & 0.044 & 0.692 & 0.045 & 0.685 & 0.042 & 0.681 \\
& WV-IC test for T2 & 0.060 & 0.671 & 0.056 & 0.663 & 0.054 & 0.688 & 0.056 & 0.690 \\
& WV-IC test for T3 & 0.040 & 0.492 & 0.049 & 0.485 & 0.044 & 0.502 & 0.037 & 0.501 \\
& WV-IC test for T4 & 0.055 & 0.507 & 0.051 & 0.497 & 0.048 & 0.502 & 0.045 & 0.497 \\
& WV-IC test for T5 & 0.044 & 0.295 & 0.049 & 0.283 & 0.046 & 0.280 & 0.053 & 0.279 \\
& WV-IC test for T6 & 0.048 & 0.299 & 0.051 & 0.290 & 0.039 & 0.289 & 0.051 & 0.291 \\
& WV-IC test for T7 & 0.031 & 0.102 & 0.036 & 0.098 & 0.034 & 0.099 & 0.049 & 0.116 \\
& WV-IC test for T8 & 0.058 & 0.118 & 0.044 & 0.118 & 0.040 & 0.107 & 0.057 & 0.104 \\
\midrule
\multicolumn{10}{l}{\textbf{Multivariate Tests (Scenario 1)}} \\
& WV-M-IC1 & 0.045 & 0.787 & 0.041 & 0.728 & 0.055 & 0.660 & 0.046 & 0.593 \\
& WV-M-IC2 & 0.047 & 0.833 & 0.043 & 0.811 & 0.035 & 0.759 & 0.037 & 0.703 \\
& Bonferroni correction & 0.052 & 0.756 & 0.052 & 0.737 & 0.037 & 0.704 & 0.040 & 0.662 \\
& Simes test & 0.053 & 0.764 & 0.052 & 0.745 & 0.038 & 0.714 & 0.044 & 0.675 \\
& ACAT & 0.055 & 0.780 & 0.050 & 0.769 & 0.040 & 0.741 & 0.051 & 0.701 \\
\midrule
\multicolumn{10}{l}{\textbf{Multivariate Tests (Scenario 2)}} \\
& WV-M-IC1 & 0.039 & 0.726 & 0.043 & 0.643 & 0.040 & 0.617 & 0.044 & 0.692 \\
& WV-M-IC2 & 0.039 & 0.781 & 0.045 & 0.753 & 0.041 & 0.677 & 0.049 & 0.588 \\
& Bonferroni correction & 0.043 & 0.709 & 0.052 & 0.698 & 0.043 & 0.650 & 0.037 & 0.604 \\
& Simes test & 0.043 & 0.722 & 0.053 & 0.717 & 0.044 & 0.662 & 0.043 & 0.619 \\
& ACAT & 0.046 & 0.743 & 0.057 & 0.722 & 0.055 & 0.694 & 0.062 & 0.651 \\
\bottomrule
\end{tabular}
\end{threeparttable}
\end{table}

\newpage 
\begin{table}[htbp]
\centering
\small
\begin{threeparttable}
\caption{Empirical size of the WV-M-IC tests under stringent p-value. Survival times are generated under the proportional hazards model (r=0).}
\begin{tabular}{llcccccccccccc}
\toprule
\multirow{3}{*}{Threshold} & \multirow{2}{*}{} 
& \multicolumn{2}{c}{\( \rho \) = 0.1} 
& \multicolumn{2}{c}{\( \rho \) = 0.25} 
& \multicolumn{2}{c}{\( \rho \) = 0.5} 
& \multicolumn{2}{c}{\( \rho \) = 0.75} \\
\cmidrule(lr){3-4} \cmidrule(lr){5-6} \cmidrule(lr){7-8} \cmidrule(lr){9-10}
& & r = 0.25 & r = 0.75 & r = 0.25 & r = 0.75 & r = 0.25 & r = 0.75 & r = 0.25 & r = 0.75 \\
\midrule
0.05 & WV-M-IC1 & 0.050 & 0.050 & 0.050 & 0.050 & 0.050 & 0.050 & 0.049 & 0.049 \\
& WV-M-IC2 & 0.049 & 0.050 & 0.050 & 0.050 & 0.051 & 0.051 & 0.048 & 0.050 \\
\midrule
0.005 & WV-M-IC1 & 0.0047 & 0.0046 & 0.0044 & 0.0049 & 0.0044 & 0.0049 & 0.0048 & 0.0051 \\
& WV-M-IC2 & 0.0048 & 0.0049 & 0.0047 & 0.0052 & 0.0049 & 0.0052 & 0.0047 & 0.0050 \\
\midrule
0.0005 & WV-M-IC1 & 0.00035 & 0.00035 & 0.00040 & 0.00054 & 0.00040 & 0.00042 & 0.00048 & 0.00048 \\
& WV-M-IC2 & 0.00037 & 0.00054 & 0.00040 & 0.00051 & 0.00048 & 0.00054 & 0.00043 & 0.00048 \\
\midrule
0.00005 & WV-M-IC1 & 0.000020 & 0.000020 & 0.000027 & 0.000033 & 0.000013 & 0.000033 & 0.000027 & 0.000073 \\
& WV-M-IC2 & 0.000033 & 0.000033 & 0.000020 & 0.000047 & 0.000033 & 0.000040 & 0.000047 & 0.000047 \\
\bottomrule
\end{tabular}
\end{threeparttable}
\end{table}

\newpage 
\begin{table}[htbp]
\centering
\small
\caption{Empirical size of the WV-M-IC tests under stringent p-value. Survival times are generated from the proportional odds model (r=1).}
\resizebox{\textwidth}{!}{%
\begin{tabular}{llcccccccccccc}
\toprule
&
& \multicolumn{2}{c}{\( \rho \) = 0.1} 
& \multicolumn{2}{c}{\( \rho \) = 0.25} 
& \multicolumn{2}{c}{\( \rho \) = 0.5} 
& \multicolumn{2}{c}{\( \rho \) = 0.75} \\
\cmidrule(lr){3-4} \cmidrule(lr){5-6} \cmidrule(lr){7-8} \cmidrule(lr){9-10}
& & r = 0.25 & r = 0.75 & r = 0.25 & r = 0.75 & r = 0.25 & r = 0.75 & r = 0.25 & r = 0.75 \\
\midrule
0.05 & WV-M-IC1 & 0.050 & 0.050 & 0.050 & 0.050 & 0.050 & 0.050 & 0.049 & 0.050 \\
& WV-M-IC2 & 0.050 & 0.050 & 0.050 & 0.050 & 0.052 & 0.051 & 0.049 & 0.051 \\
\midrule
0.005 & WV-M-IC1 & 0.0046 & 0.0043 & 0.0046 & 0.0046 & 0.0046 & 0.0046 & 0.0048 & 0.0049 \\
& WV-M-IC2 & 0.0048 & 0.0047 & 0.0044 & 0.0047 & 0.0051 & 0.0051 & 0.0046 & 0.0051 \\
\midrule
0.0005 & WV-M-IC1 & 0.00035 & 0.00029 & 0.00040 & 0.00053 & 0.00033 & 0.00042 & 0.00044 & 0.00043 \\
& WV-M-IC2 & 0.00036 & 0.00053 & 0.00039 & 0.00052 & 0.00047 & 0.00052 & 0.00036 & 0.00041 \\
\midrule
0.00005 & WV-M-IC1 & 0.000027 & 0.000020 & 0.000033 & 0.000033 & 0.000040 & 0.000040 & 0.000033 & 0.000060 \\
& WV-M-IC2 & 0.000040 & 0.000033 & 0.000020 & 0.000047 & 0.000033 & 0.000033 & 0.000060 & 0.000060 \\
\bottomrule
\end{tabular}}
\end{table}

\newpage 
\begin{table}[htbp]
\centering
\caption{Top 5 candidate genes identified by WV-M-IC1 from the ZOE 2.0 dataset under the proportional hazards model ($r=0$).}
\begin{tabular}{l ccccc}
\toprule
\textbf{} & \textit{SEPTIN14} & \textit{MORN5} & \textit{TMEM198B} & \textit{GCC2} & \textit{GTPBP3} \\
\midrule
\multicolumn{6}{l}{\textbf{Univariate Tests}} \\
\quad WV-IC test for A & $0.53234$ & $0.08110$ & $0.38291$ & $0.14742$ & $0.81451$ \\
\quad WV-IC test for B & $0.33613$ & $0.02235$ & $0.91147$ & $0.04073$ & $0.39359$ \\
\quad WV-IC test for I & $0.31113$ & $0.01488$ & $0.24742$ & $0.02149$ & $0.74301$ \\
\quad WV-IC test for J & $0.28145$ & $0.00201$ & $0.15511$ & $0.24497$ & $0.59156$ \\
\quad WV-IC test for K & $0.69506$ & $0.19552$ & $0.41778$ & $0.00441$ & $0.36856$ \\
\quad WV-IC test for L & $0.54302$ & $0.00843$ & $0.03716$ & $0.31749$ & $0.75855$ \\
\quad WV-IC test for S & $0.19634$ & $0.00012$ & $0.00018$ & $0.00493$ & $0.92274$ \\
\quad WV-IC test for T & $0.26822$ & $0.01960$ & $0.43895$ & $0.20499$ & $0.00165$ \\
\midrule
\multicolumn{6}{l}{\textbf{Multivariate Tests}} \\
\quad Bonferroni        & $1.00000$ & $0.00094$ & $0.00140$ & $0.03525$ & $0.01323$ \\
\quad Simes             & $0.53781$ & $0.00094$ & $0.00140$ & $0.01971$ & $0.01323$ \\
\quad ACAT              & $0.37139$ & $0.00086$ & $0.00139$ & $0.01548$ & $0.01373$ \\
\quad WV-M-IC1          & $0.00013$ & $0.00024$ & $0.00032$ & $0.00033$ & $0.00044$ \\
\quad WV-M-IC2          & $0.37250$ & $0.00133$ & $0.05171$ & $0.01503$ & $0.23433$ \\
\midrule
\quad $p$-value threshold & $4.05\times 10^{-7}$ & $8.11\times 10^{-7}$ & $1.22\times 10^{-6}$ & $1.62\times 10^{-6}$ & $2.03\times 10^{-6}$ \\
\bottomrule
\end{tabular}
\end{table}

\newpage 
\begin{table}[htbp]
\centering
\caption{Top 5 genes identified by WV-M-IC2 from the ZOE 2.0 dataset under the proportional hazards model ($r=0$).}
\begin{tabular}{l ccccc}
\toprule
\textbf{} & \textit{MAN1A2} & \textit{LINC01376} & \textit{LOC105377267} & \textit{KRTAP3-3} & \textit{FAM234A} \\
\midrule
\multicolumn{6}{l}{\textbf{Univariate Tests}} \\
\quad WV-IC test for A & $6.94\times 10^{-5}$ & $0.00023$ & $0.00923$ & $0.01201$ & $0.00115$ \\
\quad WV-IC test for B & $0.00698$ & $0.00772$ & $0.00702$ & $0.00051$ & $0.00457$ \\
\quad WV-IC test for I & $0.00062$ & $0.00038$ & $0.01441$ & $0.00720$ & $0.00054$ \\
\quad WV-IC test for J & $0.00073$ & $0.00011$ & $0.01169$ & $0.04555$ & $0.00350$ \\
\quad WV-IC test for K & $9.70\times 10^{-5}$ & $0.00223$ & $0.00051$ & $0.00269$ & $0.00902$ \\
\quad WV-IC test for L & $0.00724$ & $0.00016$ & $0.00072$ & $0.00033$ & $0.01940$ \\
\quad WV-IC test for S & $0.00293$ & $0.00195$ & $0.00013$ & $0.00012$ & $0.00047$ \\
\quad WV-IC test for T & $0.01758$ & $0.02895$ & $0.00131$ & $0.00219$ & $0.00312$ \\
\midrule
\multicolumn{6}{l}{\textbf{Multivariate Tests}} \\
\quad Bonferroni        & $0.00056$ & $0.00088$ & $0.00107$ & $0.00099$ & $0.00375$ \\
\quad Simes             & $0.00039$ & $0.00062$ & $0.00107$ & $0.00099$ & $0.00216$ \\
\quad ACAT              & $0.00028$ & $0.00034$ & $0.00067$ & $0.00057$ & $0.00137$ \\
\quad WV-M-IC1          & $0.01921$ & $0.00214$ & $0.01297$ & $0.02549$ & $0.02629$ \\
\quad WV-M-IC2          & $3.12\times 10^{-5}$ & $3.77\times 10^{-5}$ & $0.00011$ & $0.00013$ & $0.00017$ \\
\midrule
\quad $p$-value threshold & $4.05\times 10^{-7}$ & $8.11\times 10^{-7}$ & $1.22\times 10^{-6}$ & $1.62\times 10^{-6}$ & $2.03\times 10^{-6}$ \\
\bottomrule
\end{tabular}
\end{table}

\newpage 
\begin{table}[htbp]
\centering
\caption{Top 5 candidate genes identified by WV-M-IC1 from the ZOE 2.0 dataset under the proportional Odds model ($r=1$).}
\begin{tabular}{l ccccc}
\toprule
\textbf{} & \textit{SEPTIN14} & \textit{MORN5} & \textit{GTPBP3} & \textit{ZNF587} & \textit{GCC2} \\
\midrule
\multicolumn{6}{l}{\textbf{Univariate Tests}} \\
\quad WV-IC test for A & $0.53156$ & $0.05949$ & $0.78212$ & $0.41281$ & $0.15253$ \\
\quad WV-IC test for B & $0.35429$ & $0.01988$ & $0.36113$ & $0.00151$ & $0.04268$ \\
\quad WV-IC test for I & $0.30802$ & $0.01213$ & $0.71382$ & $0.87754$ & $0.01839$ \\
\quad WV-IC test for J & $0.25856$ & $0.00254$ & $0.59027$ & $0.22404$ & $0.22329$ \\
\quad WV-IC test for K & $0.64926$ & $0.18627$ & $0.34530$ & $0.78494$ & $0.00509$ \\
\quad WV-IC test for L & $0.51511$ & $0.00508$ & $0.76700$ & $0.14932$ & $0.35136$ \\
\quad WV-IC test for S & $0.15442$ & $8.58\times 10^{-5}$ & $0.91800$ & $0.17124$ & $0.00826$ \\
\quad WV-IC test for T & $0.24834$ & $0.01660$ & $0.00104$ & $0.90555$ & $0.21426$ \\
\midrule
\multicolumn{6}{l}{\textbf{Multivariate Tests}} \\
\quad Bonferroni        & $1.00000$ & $0.00069$ & $0.00836$ & $0.01208$ & $0.04072$ \\
\quad Simes             & $0.56686$ & $0.00069$ & $0.00836$ & $0.01208$ & $0.03306$ \\
\quad ACAT              & $0.34046$ & $0.00064$ & $0.00853$ & $0.01219$ & $0.01947$ \\
\quad WV-M-IC1          & $7.26\times 10^{-5}$ & $0.00018$ & $0.00031$ & $0.00034$ & $0.00038$ \\
\quad WV-M-IC2          & $0.33333$ & $0.00097$ & $0.21571$ & $0.17969$ & $0.01725$ \\
\midrule
\quad $p$-value threshold & $4.05\times 10^{-7}$ & $8.11\times 10^{-7}$ & $1.22\times 10^{-6}$ & $1.62\times 10^{-6}$ & $2.03\times 10^{-6}$ \\
\bottomrule
\end{tabular}
\end{table}

\newpage 
\begin{table}[htbp]
\centering
\caption{Top 5 genes identified by WV-M-IC2 from the ZOE 2.0 dataset under the proportional Odds model ($r=1$).}
\begin{tabular}{l ccccc}
\toprule
\textbf{} & \textit{MAN1A2} & \textit{LINC01376} & \textit{LOC105377267} & \textit{MIR3123} & \textit{KRTAP3-3} \\
\midrule
\multicolumn{6}{l}{\textbf{Univariate Tests}} \\
\quad WV-IC test for A & $6.97\times 10^{-5}$ & $0.00025$ & $0.01251$ & $0.00163$ & $0.01457$ \\
\quad WV-IC test for B & $0.00645$ & $0.00678$ & $0.00703$ & $0.04103$ & $0.00046$ \\
\quad WV-IC test for I & $0.00034$ & $0.00034$ & $0.01376$ & $0.00361$ & $0.00713$ \\
\quad WV-IC test for J & $0.00077$ & $0.00014$ & $0.01366$ & $0.00043$ & $0.05357$ \\
\quad WV-IC test for K & $6.41\times 10^{-5}$ & $0.00208$ & $0.00047$ & $0.00844$ & $0.00529$ \\
\quad WV-IC test for L & $0.00535$ & $0.00015$ & $0.00068$ & $0.00279$ & $0.00031$ \\
\quad WV-IC test for S & $0.00220$ & $0.00166$ & $0.00012$ & $3.45\times 10^{-5}$ & $9.88\times 10^{-5}$ \\
\quad WV-IC test for T & $0.01869$ & $0.02942$ & $0.00119$ & $0.01173$ & $0.00263$ \\
\midrule
\multicolumn{6}{l}{\textbf{Multivariate Tests}} \\
\quad Bonferroni        & $0.00051$ & $0.00111$ & $0.00094$ & $0.00028$ & $0.00079$ \\
\quad Simes             & $0.00028$ & $0.00059$ & $0.00094$ & $0.00028$ & $0.00079$ \\
\quad ACAT              & $0.00023$ & $0.00036$ & $0.00060$ & $0.00024$ & $0.00049$ \\
\quad WV-M-IC1          & $0.01182$ & $0.00196$ & $0.01016$ & $0.01312$ & $0.02212$ \\
\quad WV-M-IC2          & $2.16\times 10^{-5}$ & $3.42\times 10^{-5}$ & $0.00012$ & $0.00013$ & $0.00015$ \\
\midrule
\quad $p$-value threshold & $4.05\times 10^{-7}$ & $8.11\times 10^{-7}$ & $1.22\times 10^{-6}$ & $1.62\times 10^{-6}$ & $2.03\times 10^{-6}$ \\
\bottomrule
\end{tabular}
\end{table}


\end{document}